\documentclass[journal]{IEEEtran}

\usepackage{times}

\usepackage{graphicx}

\usepackage{amsmath}
\usepackage{times}
\usepackage{amsfonts}
\usepackage{amssymb}
\usepackage{cite}
\usepackage{caption}
\usepackage{graphicx} 
\usepackage[caption=false, font=normalsize, labelfont=sf, textfont=sf]{subfig}
\usepackage{url}
\usepackage[dvipsnames]{xcolor}
\usepackage{color, colortbl}
\usepackage{caption}
\usepackage{subcaption}
\usepackage{multirow}
\usepackage{stfloats} 
\usepackage[numbers,sort&compress]{natbib} 
\usepackage{lipsum}
\usepackage[caption=false, font=normalsize, labelfont=sf, textfont=sf]{subfig}

\usepackage[ruled,vlined]{algorithm2e}

\DeclareGraphicsExtensions{.png,.eps,.ps,.pdf}

\begin{document}

\title{Exploring Indoor Localization for Smart Education} 

\author{
\IEEEauthorblockN{Carlos S. Álvarez-Merino\IEEEauthorrefmark{1},
Emil J. Khatib\IEEEauthorrefmark{1},
Antonio Tarrías Muñoz\IEEEauthorrefmark{1} and 
Raquel Barco\IEEEauthorrefmark{1} }
\\
\IEEEauthorblockA{\IEEEauthorrefmark{1}Telecommunication  Research  Institute  (TELMA),  Universidad  de  Málaga,  E.T.S.I. de Telecomunicación, Bulevar Louis Pasteur 35, 29010, Málaga, Spain\\ Email: cam@ic.uma.es, emil@uma.es, atm@ic.uma.es,  rbarco@uma.es} 

\thanks{This work has been performed in the framework of the MAORI project (grant agreement number TSI-063000-2021-53) funded by the European Union-\textit{NextGenerationEU}}
}

\maketitle

\begin{abstract}                          
This comprehensive study delves into the realm of indoor positioning technologies within the domain of Smart Education (SE). Focusing on typical techniques and technologies in educational settings, the research emphasizes the importance and potential services of localization in SE. Moreover, this work explores the feasibility and limitations of these technologies, providing a detailed account of their role in educational settings. The paper also contains in an innovative Proof of Concept (PoC), demonstrating an automatic attendance control (AAC) system that integrates 5G and WiFi technologies. This PoC effectively showcases the possibilities and effectiveness of location-based services in educational surroundings even with a limited budget, setting the stage for optimizing teaching time, enhancing the quality of education.

\end{abstract}

\begin{IEEEkeywords}                           
Smart Education, localization, 5G, WiFi FTM, fusion, ML, Random Forest

\end{IEEEkeywords}                             

\section{Introduction} \label{INT} 

In the last years, the impact of precise location services is growing in society with the advancements in technology, leading to the use of location-based services such as autonomous robots, e-Health or context-aware applications to provide personalized services \cite{khatib2023designing}. By providing more accurate and reliable location information, these services can help improve safety, efficiency, and overall effectiveness in a wide range of industries and applications. Location-based services have also had a transformative impact on Smart Education (SE) \cite{gros2016design}, where they can be used to improve the overall efficiency and effectiveness by providing real-time information about the location of resources \cite{sayapogu2021ar}, such as classrooms and labs, and the location of students or teachers.

Students are now more connected and engaged when carrying out their academic activities with digital devices. Given the increasing inclination of students to work digitally, the educational infrastructure must satisfy this digital offer towards better future generations. SE consists on the use of the technology to optimize the whole educational system for a personalized teaching and learning process for each student \cite{zhu2016research,chengoden2023metaverse}. This includes the use of advanced algorithms, data analytics, and Artificial Intelligence (AI) to create flexible and interactive learning environments that can adapt to the unique needs, preferences, and abilities of each learner \cite{liu2019multi}. SE utilizes a diverse range of digital tools and platforms, including mobile devices, learning management systems, educational apps, and Virtual Reality (VR) technologies, to develop captivating learning environments \cite{sungkur2016augmented}.


In the context of SE, the need for accurate location estimation is crucial for personalized services. The most common approach for precise localization is the use of Global Navigation Satellite System (GNSS), which provides high accuracy in outdoor scenarios. However, GNSS is not available indoors, where many applications for this field are being developed, due to signal blocking or signal reflections. To overcome this limitation, various technologies and techniques such as 5G and WiFi are being used to provide accurate and precise location information indoors and in built-up areas \cite{alvarez2021wifi}. Additionally, some applications require the network to estimate the location of end-users in order to save energy and reduce computational complexity \cite{sciarrone2016smart}. Network-based location is a better solution for these functions, as it allows the network to estimate the location of terminals based on data collected in the network infrastructure without requiring cooperation from the terminals.




This paper provides an overview of the role of localization within the services existing in an SE vertical scenario while also offering insights into potential forthcoming services in the upcoming years. In addition, this work contributes to the development of an Automatic Attendance Control (AAC) system that is carried out as a Proof of Concept (PoC) to demonstrate a location-based service within a resource-constrained SE.

\begin{figure}[!h]
\centering
\includegraphics[width=1.0\columnwidth]{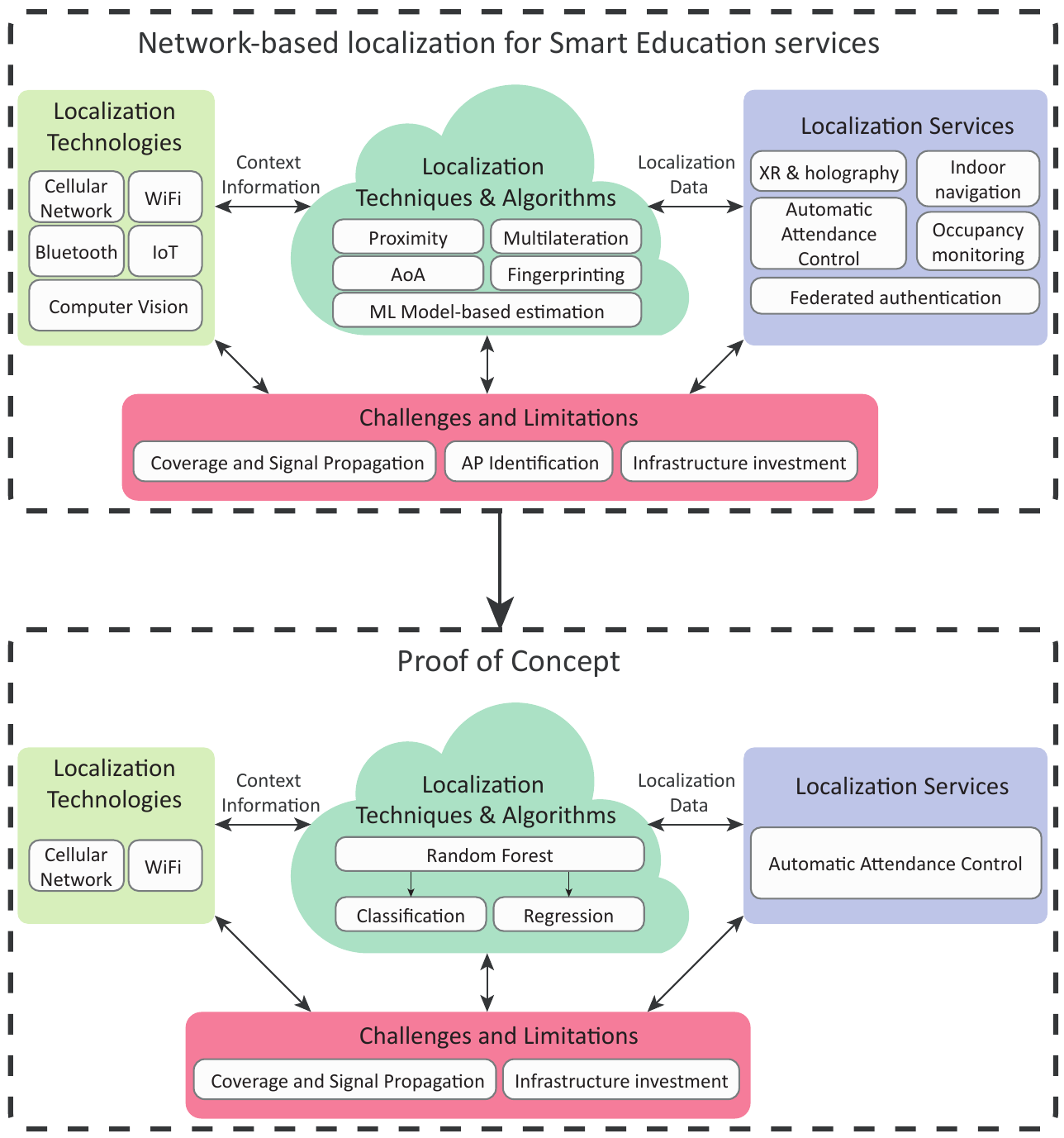}
\caption{Structure of the paper}
\label{Schema}
\end{figure}

The rest of the paper is organized as follows. Section \ref{LOC} provides a comprehensive analysis of various use cases for location-based services in SE. It delves into the requirements and challenges of localization associated with these services.
In Section \ref{TEC}, various localization techniques commonly applied in the educational settings are explored. Section \ref{TECH} delves into localization technologies that are in the context of SE. In Section \ref{PRIV}, privacy concerns of localization within an educational context are explained. Section \ref{POC} explains the PoC, highlighting its objective of AAC through Machine Learning (ML) and detailing the methodology as well as employed classification and regression models. Additionally, the experimental setup and the results are presented and discussed. Finally, the conclusions are carefully reviewed in Section \ref{CONC}. Figure \ref{Schema} presents the structure of the paper.
The acronyms in this paper are listed in the Table \ref{Table_Acro}.

\begin{table}[h]
\captionsetup{justification=centering}
\caption{Overview of acronyms}
\centering
\begin{tabular}{cc}
\hline \hline
\textbf{Acronym}    & \textbf{Definition}                             \\ \hline \hline
AoA           & Angle of Arrival                                      \\ \hline
AP            & Access Point                                          \\ \hline
ANN           & Approximate Nearest Neighbors                         \\ \hline
AI            & Artificial Intelligence                               \\ \hline
AR            & Augmented Reality                                     \\ \hline
BLE           & Bluetooth Low Energy                                  \\ \hline
CSI           & Channel State Information                             \\ \hline
CDF           & Cumulative Density Function                           \\ \hline
DFL           & Device-Free Localization                              \\ \hline
EU            & European Union                                        \\ \hline
FR            & Frequency Range                                       \\ \hline
GDPR          & General Data Protection Regulation                    \\ \hline
GNSS          & Global Navigation Satellite System                    \\ \hline
gNB           & gNodeB                                                \\ \hline
IMU           & Inertial Measurement Unit                             \\ \hline
IoT           & Internet of Things                                    \\ \hline
LOS           & Line-of-Sight                                         \\ \hline
LS            & Least Squares                                         \\ \hline
MIMO          & Multiple Input Multiple Output                        \\ \hline
ML            & Machine Learning                                      \\ \hline
MR            & Mixed Reality                                         \\ \hline
NB            & Narrow Band                                           \\ \hline
NRPPa         & New Radio Positioning Protocol A                      \\ \hline
PAN           & Personal Area Network                                 \\ \hline
PIR           & Passive Infra-Red                                     \\ \hline
PoC           & Proof of Concept                                      \\ \hline
pRRH          & pico-Remote Radio Heads                               \\ \hline
RF            & Random Forest                                         \\ \hline
RSSI          & Received Signal Strength Indicator                    \\ \hline
RTT           & Round Trip Time                                       \\ \hline
SE            & Smart Education                                       \\ \hline
SSID          & Service Set Identifier                                \\ \hline
ToF           & Time of Flight                                        \\ \hline
UWB           & Ultra Wide Band                                       \\ \hline
UE            & User Equipment                                        \\ \hline
VR            & Virtual Reality                                       \\ \hline
XR            & eXtended Reality                                      \\ \hline
3GPP          & 3rd Generation Partnership Project                    \\ \hline

\end{tabular}
\label{Table_Acro}
\end{table}

\section{Localization Services in Smart Education} \label{LOC} 


Education is one of the key pillars of modern society. As human knowledge advances, the topics that are taught become more and more complex and profound, and the teaching methods must evolve and adapt to new layers of complexity \cite{fortes2019campus,zhu2016research}. For this reason, education is a very dynamic market, that adopts not only new teaching methods, but also new technologies. Localization, in particular, introduces a diverse range of services aimed at actively involving and inspiring students. In this section, an overview of different localization-based applications in SE are presented. 






\subsection{XR and holography}

eXtended Reality (XR) is a comprehensive term encompassing a range of immersive technologies that merge both digital and physical worlds. It includes VR, Augmented Reality (AR) and Mixed Reality (MR), enabling users to simultaneously immerse themselves in and interact with virtual and real environments \cite{kosko2021conceptualizing}. On the other hand, holography refers to the technique of encoding a light field as an interference pattern of phase and amplitude variations. When appropriately illuminated, a hologram diffracts incident light, creating a faithful replica of the initial light field, resulting into a realistic representation of the recorded 3D objects \cite{shi2021towards}. Both technologies enable an immersive experience that transcends the boundaries of conventional media, offering unique opportunities for diverse applications in fields such as entertainment, education, healthcare, and engineering.

Within the realm of SE, the utilization of techniques like gamification, which involves converting educational concepts into game-like formats and leveraging the brain's dopamine response to improve the learning experience, presents an opportunity to captivate students more effectively and to foster student engagement \cite{lumsden2016gamification,jagoda2020experimental}. Such gamification strategies heavily rely on advanced technologies such as XR or holography \cite{imoize20216g}, which impose substantial demands on processing power and communication capabilities, while also emphasizing the need for physical portability and non-intrusiveness. 

By incorporating gesture recognition \cite{saggio2018sensory} and location \cite{chengoden2023metaverse}, interaction with AR/VR objects can be facilitated, thereby simplifying the complexity and cost of the end devices \cite{eswaran2022challenges}. Furthermore, location serves as a crucial factor in the traffic generated by SE applications, enabling efficient network management \cite{alexandropoulos2021hybrid}. XR requires a latency below 50 ms \cite{alriksson2021xr} and achieves a localization accuracy of 0.1 meters \cite{carmigniani2011augmented}. On the other hand, holography can experience a latency as high as 100 ms \cite{elzanaty2023towards, hooft2019tile} while localization accuracy must be below the centimeter-level \cite{elzanaty2023towards}. When students are situated within a classroom, broadband traffic becomes concentrated in a hotspot. This traffic is similar among students but varies slightly based on their precise location, such as different viewing angles of the same XR object. Hence, if the location is known, the usage of edge resources can be optimized \cite{hooft2019tile}.

In the next decade, XR and holography could also be used to create virtual classrooms, where students can attend classes remotely and interact with teachers and classmates in real-time. This could open up new possibilities for education, such as providing access to education to remote and underserved communities \cite{fox2022white}.

\subsection{Indoor navigation}

Indoor navigation is a technology that can be used to provide positioning, guidance and wayfinding for people within a building or campus \cite{ng2020design}. To provide a reliable and seamless indoor navigation service, localization accuracy must be enhanced to the level of meter-level in horizontal plane \cite{zafari2019survey} and floor-level in height \cite{el2021indoor}. This level of precision is necessary to ensure that users can confidently navigate within indoor spaces, avoiding obstacles and reaching their intended destinations accurately. However, by leveraging a combination of position and inertial sensors, including Inertial Measurement Unit (IMU), it becomes possible to achieve a comprehensive and accurate navigation experience even when the localization precision is diminished \cite{malyavej2013indoor}.

Different technologies have claimed to provide precise localization down to some meter-level or lower. For instance, Ultra Wide Band (UWB) technology achieves centimeter-level accuracy with time-based estimations \cite{tian2020research} meanwhile WiFi 802.11mc obtains an accuracy of 1-2 meters with the same protocol \cite{gentner2020wifi}. 5G claims to provide $<$3 meters for 80\% of the cases, encompassing both horizontal and vertical planes \cite{location3gpp} with the aim of indoor navigation. Bluetooth Low Energy (BLE) fulfills the requirements of precision for indoor navigation when mapping the whole scenario in a previous step for fingerprinting \cite{aranda2022performance}. However, fingerprinting is not practical for precise navigation around a whole campus due to the cost of deployment \cite{rocamora2020survey}. While 5G or WiFi Access Points (APs) would provide services, mainly internet access, as well as localization features, a BLE-based deployment would not offer additional functionalities except for localization. Furthermore, for indoor navigation systems to be truly effective, a real-time location service is required. Users require instant updates and guidance to make informed decisions while traversing indoor environments. Therefore, the system should operate with a maximum latency of 1 s \cite{soyer2021efficient}, minimizing any perceptible delays for a fluent navigation response \cite{zafari2019survey}.

Indoor navigation can be used to improve the overall efficiency and experience of SE by reducing congestion and making it easier for people to find their way to the desired classroom or laboratory \cite{ng2020design}. Moreover, it can set up virtual boundaries, a feature called geo-fencing that triggers an action when a device or a person enters or exits that boundary \cite{joshi2022gnl}. Geo-fencing is a powerful tool that is often used in security systems to restrict access to certain areas and ensure that only authorized personnel are present in sensitive areas. 

Indoor navigation is also used in emergency situations such as fires, earthquakes, or other disasters where time is crucial, and having a reliable indoor navigation system can be critical to save lives. It can also be integrated with other emergency systems, such as fire alarms, smoke detectors, and emergency lighting, to provide a comprehensive solution for emergency preparedness \cite{wong2023indoor}. In the event of an emergency, the system would automatically trigger an alert, providing immediate guidance to individuals in the affected area making it easier for people to quickly and safely evacuate the building while avoiding certain spaces \cite{anthony2019universally}.

\subsection{Occupancy monitoring}

Occupancy monitoring is an important issue in SE, as overcrowding can lead to safety concerns, reduced comfort and productivity, and increased wear and tear on facilities \cite{singh2020smart, azizi2020application}. It allows administrators to track the occupancy of different areas in real-time and enforce safe capacity limits. In addition, real-time occupancy monitoring can provide valuable insights into how different areas of the campus are being used, allowing administrators to optimize the use of resources and energy efficiency by combining with different actuators such as lighting or air conditioners, e.g. identifying areas of low occupancy and adjusting the lighting and temperature accordingly. 

To accomplish this task, Device-Free Localization (DFL) is a technology that can be used to track the presence and movement of people in a given environment without the need for them to carry any device. There are two types of DFL: based on images and signal propagation. 

Camera or vision-based systems combined with ML algorithms provide a centimeter-level accuracy \cite{morar2020comprehensive}; however, partial occlusion results in coverage blind spots and privacy concerns make this method unfeasible for urban areas \cite{ngamakeur2020survey}. Alternatively, DFL systems based on signal propagation typically rely on WiFi, Zigbee and UWB technologies \cite{ngamakeur2020survey}. When utilizing Channel State Information (CSI), DFL systems capture the multipath propagation during the wireless transmission offering a nonintrusive approach with high sensitivity to channel variations \cite{xue2022enhanced}. These inherent characteristics make CSI-based DFL systems within an horizontal error of few meters \cite{basiri2017indoor}. Moreover, DFL systems offer a latency below 1 s \cite{ngamakeur2020survey}. DFL can also be based on the Received Signal Strength Indicator (RSSI) between a transmitter and a receiver with LoS \cite{sasiwat2019experimental}. Given that the human body consists of approximately 70\% water, it absorbs radio signals, leading to shadowing effects \cite{kolahi2019impact}. This process primarily focuses on human movement and tracking \cite{abdull2020rssi, konings2019springloc}. Signal propagation-based DFL systems can be categorized as either model-based \cite{wang2018wi} or fingerprint-based methods \cite{mager2015fingerprint}. 

There are several less effective technologies for nonintrusive DFL, including air-pressure sensing and ultrasound signal reflections \cite{hazas2002novel}, Passive Infra-Red (PIR) which detects thermal energy radiation from the human body \cite{delaney2009evaluation}, and CO2 concentration measurement in buildings \cite{leephakpreeda2001occupancy}.


\subsection{Automatic Attendance Control}


Early attempts at attendance control was a labor-intensive and time-consuming process, and the accuracy and reliability of the attendance data could vary depending on the expertise of the educators. Attendance control is critical for learning, ensuring that students are participating in classes and receiving the education they need to succeed \cite{bartanen2020principal}. 

Over time, advancements in technology have made localization in education more efficient and effective. Up to date, the use of AI and ML has allowed institutions to automatically translate the geo-location of a teacher to attendance control of its workplace when they are in the nearby of the institution \cite{yousef2021auto}. Thus, the rise of SE have evolved to encompass the need for accurate localization data to track students' attendance during lectures \cite{abdel2019internet}. 

\begin{table*}[!b]
\centering
\caption{Comparison of different use cases}
\resizebox{\textwidth}{!}
{
\begin{tabular}{cccc}
\hline \hline
Use Case                   & Location Accuracy & Latency & Description                                     \\ \hline \hline
XR                         & 0.1 m             & 50 ms   & \multirow{2}{*}{Immersive and interactive systems to improve the learning experience} \\ \cline{1-3}
Holography                 & 1 cm              & 100 ms  &                                                 \\ \hline
Indoor Navigation          & 3 m               & 1 s     & Guidance for finding people/offices within a building or campus  \\ \hline
\begin{tabular}[c]{@{}c@{}}Space Usage and \\ Ocuppancy Tracking\end{tabular}  & 10 m              & 1 s     & Capacity control with DFL systems for safety concerns           \\ \hline
Attendance Control         & 5 m             & 1 s     & To detect the presence whether a student is in a classroom     \\ \hline
\end{tabular}
}
\label{TableComparison}
\end{table*}

There are different solutions to control the attendance of the students. Camera-based systems identifies with high probability the face of the student to track their attendance \cite{strueva2021student}. However, this type of system carries significant  privacy concerns with additional issue of the subjects often being underage \cite{qureshi2021internet}. 
Barcode or QR scan \cite{mustafa2022electronic} or RFID identification systems \cite{ula2021new} solves these privacy concerns. In barcode or QR systems, students must log in through the institute portal by an application to ensure their attendance by face id \cite{singh2019attendance} or biometrics \cite{varadannanavar2022automatic}. RFID identification systems check the attendance based on an NFC system that detects the students at the beginning of the class. QR scanning systems are more time efficient because it makes all students to ensure their attendances in a time interval. Nevertheless, this process halts the lecture for a brief duration. 

Automatic attendance control eliminates these processes on the user's side to reduce the time consumed in this process to zero. Moreover, it aids educators in identifying and addressing any issues or challenges that students may be facing, such as absenteeism or lack of engagement. Additionally, attendance data can be used to evaluate the effectiveness of educational programs and make improvements where needed. In short, attendance control is an essential aspect of ensuring that students are receiving a high-quality education. 

In \cite{puckdeevongs2020classroom}, an automatic attendance control system based on localization is developed based on BLE. This system matches the localization of the students during the lecture with the attendance control. During this process, neither students nor teacher intervene during the process of attendance control. For this process, time is not critical, so even a latency of up to a few seconds is valid \cite{allen2017authenticating}. The location process is done with Approximate Nearest Neighbors (ANN), which is an ML model, that estimates the position of the user by a previous modelling of the scenario. Radio technologies that are usually present in SE scenarios, such as 5G, WiFi or BLE, can also be leveraged for this use. By using the RSSI information, the system can automatically detect the presence of a student in a classroom within a localization accuracy of 2-5 meters \cite{ji2021multivariable,wang2018wifi}.


Table \ref{TableComparison} shows an overview of the different use cases with their minimum location accuracy and latency required for the 80\% of the cases for 5G commercial use cases \cite{location3gpp} and a brief description of their use in SE.

\section{Localization Techniques} \label{TEC} 

Depending on the nature of the data information, e.g. signal power, distance or angle to the AP, different approaches are commonly utilized to locate users, including location by proximity, ranging-based methods, Angle of Arrival (AoA), fingerprinting and model-based localization.

Location by proximity is the simplest method for determining the User Equipment's (UE) location, assuming it to be the same as the AP location. This method is employed when high accuracy is not a strict demand \cite{satan2018development}.

Ranging-based techniques, such as multilateration, involve computing distances to APs using metrics such as RSSI or Time of Flight (ToF) \cite{alvarez2021wifi}. These methods achieve a high accuracy if the ranges are precise. The final location estimation is determined by the intersection of spheres (or circles in 2D). Nevertheless, range estimations are prone to result into non-convergence of the circles or hyperbolas used in the trilateration process. To mitigate this uncertainty, techniques such as Least Squares (LS) are applied \cite{teunissen2017springer}.

AoA measures the angle at which the signal arrives at the UE from the AP. This approach is employed in Multiple Input Multiple Output (MIMO) systems because of their ability to utilize beamforming techniques \cite{zhu2017auxiliary}. Indoor environments present challenges for both range-based models and AoA due to signal blocking and reflections \cite{liu2020survey}.

In cases where received power remains relatively constant over time, despite not following a predetermined propagation model, it can serve as a stable reference. For instance, in a location close to an AP, if the measured power is consistently reduced due to an obstacle like a wall, this power level remains constant as long as the obstruction remains unchanged. Each point in space is associated with paired values comprising reference point identifiers and unchanging received power levels, forming a unique signature known as a \textit{fingerprint} \cite{lee2020precise}. However, fingerprinting has notable limitations, including sensitivity to variations in training and testing conditions caused by dynamic propagation attributes like temperature, humidity, and obstacles \cite{maneerat2019roc, ye2020wireless, hayward2022survey}. It also requires an initial radio map construction phase that limits the covered area, unrecorded data points cannot be used for positioning during operational phases \cite{kawecki2022performance}. 

To address this issue, a commonly used approach is to employ ML algorithms to create an environment model \cite{maduranga2021treeloc, maduranga2021supervised, janssen2019comparing}, which can then be utilized for estimating the position during the exploitation phase. ML algorithms generate a comprehensive model of the scenario through the information provided in the training phase with a reduced number of \textit{fingerprints}. Consequently, the ML models enable a localization service encompassing the entire dessignated area \cite{rathnayake2023rssi, alhajri2019indoor}.

\section{Localization Technologies for SE} \label{TECH} 

This section explores the radio technologies that are commonly present in educational environments. These technologies are often installed for network access, serving as a backbone for SE services. The purpose of this paper is to present them and define how they can also be opportunistically used for indoor positioning.

\subsection{Cellular Network}

Cellular networks offer a myriad of services based on voice and data traffic. Currently, this technology is prevalent in educational institutions as it enables us to access any information source at any time. Many infrastructures are deployed in educational institutions, by operators to handle high densification, and experimental networks by certain universities for research and development purposes.

Although we are at an early stage of 5G deployment, the 3rd Generation Partnership Project (3GPP) has formally declared its commitment to achieving an accuracy of less than 3 meters in both horizontal and vertical dimensions, and up to 10 meters in the vertical plane in open spaces for 80\% of the cases 3GPP. To this end, various protocols and techniques will be employed, including the deployment of the multi-Round Trip Time (RTT) protocol \cite{dwivedi2021positioning}, which uses timestamps to measure the distance between the UE and the different cells to improve the accuracy of the system. The 3GPP has noted in technical reports, such as \cite{location3gpp}, that the use of RTT can be effectively employed in both frequency spectrums defined within the 5G framework. These spectrums encompass Frequency Range (FR) 1 for frequencies below 6GHz and FR 2 designed for the millimeter band (mmWave). This protocol will be used in both upstream and downstream communication channel, not only from the serving cell but also from neighboring cells. This approach is aimed of obtaining precise location of users without incurring higher energy costs.

The implementation of 5G operating at high frequencies presents new technical challenges in comparison to lower and mid-band services \cite{kim2020coverage}. Initially, mmWaves have a centimeter-level location precision \cite{yammine2021experimental} but a shorter propagation distance, resulting in greater Line-of-Sight (LoS) path loss than sub-6GHz waves, necessitating smaller cell sizes and/or more powerful radio stations \cite{TR38.901}. Additionally, mmWaves do not propagate through many of the external/internal building materials such as concrete walls \cite{haneda2016indoor,du2018suburban}. Despite of being covered, certain fluctuations or areas devoid of connectivity can arise challenges that result in the inconsistent availability of SE services.

It is important to consider the evolving landscape of mobile networks, which are increasingly moving towards the deployment of smaller cellular units. These smaller cells are expected to be highly integrated into SE \cite{fortes2019campus}. The adoption of 
advanced technologies like beamforming \cite{lazarev2019positioning} and the densification of cells will allow the presented services in SE. Further network densification results in UEs' receiving greater contextual information, such as distances/angles between APs and UEs, that substantially improves location services, among others. Nonetheless, some cellular network manufacturers have opted for pico-Remote Radio Heads (pRRH) -based infrastructure in industrial deployments. This means that there exist different 5G APs that are operating as a single cell, in a synchronized way. This approach provides several benefits such as the avoidance of handovers between APs when the users are moving. Nevertheless, it makes more difficult to use the 5G APs for localization, since it is not possible to distinguish which AP the user is connected to \cite{sigwele2018intelligent}.

\subsection{WiFi}

Due to its widespread availability and worldwide deployment, WiFi networks offer global coverage in educational areas that can be the backbone for location-based services within SE. Eduroam is the secure, worldwide roaming access educational network developed for the international research and education, accessible to all students \cite{fiore2023eduroam}. This network, based on a single shared SSID (Service Set Identifier), has become a standard that selects APs and enables roaming, guaranteeing continuous connectivity while moving between campuses or affiliated institutions \cite{florio2005eduroam}. Eduroam stands out for providing a secure worldwide service through the implementation of robust authentication protocols like EAP and WPA2-Enterprise encryption \cite{perkovic2020wpa2}.

When WiFi employs the IEEE 802.11mc standard, it incorporates a feature known as Fine Time Measurement (FTM), which facilitates precise distance estimation from the UE to the AP. This estimation is accomplished by the insertion of timestamps and the utilization of the RTT protocol \cite{hashem2021accurate,gentner2020wifi}. This release is intended to transform the indoor positioning industry in the coming years, as new smartphones are adopting the IEEE 802.11mc protocol universally \cite{WifiFtmDeveloper}. Implementing a 5G network and the necessary infrastructure can be expensive \cite{wisely2018capacity}, and may not be feasible for all schools or educational institutions. Thus, WiFi technology with eduroam network only needs to change the APs to implement the IEEE 802.11mc protocol to provide an accurate localization service. 

The protocol estimates with an accuracy of around one meter the distance of any user that supports the protocol without the need to be connected to the AP \cite{martin2020passive}. The information is calculated on the UE side to safeguard user privacy, since the location information is not shared among the nodes in the network. Nevertheless, it is also remarkable the extensive study of the conventional WiFi for localization using signal power \cite{liu2011wifi, bisio2016smart,liu2020survey} known as \textit{fingerprinting}.

\subsection{Bluetooth}

Bluetooth is an ubiquitous technology owing to its widespread adoption in Personal Area Networks (PANs) such as smartwatches, headphones or smartphones. It is a short-range wireless communications technology which facilitates cost-effective, low-bandwidth, and energy-efficient communication thanks to the BLE protocol \cite{collotta2017innovative}. Notably, BLE-based positioning relies on the measurement of signal power as a key determinant \cite{li2018indoor}. However, this technology offers low precision in localization terms. Under specific LoS conditions and with proximity to APs, an accuracy of only a few meters error can be obtained \cite{kluge2020trilateration}. Furthermore, BLE holds considerable promise for forthcoming sensor implementations in the Internet of Things (IoT) \cite{pallavi2019overview,jeon2018ble}. 
Consequently, while Bluetooth technology demonstrates considerable potential, it is not yet as implemented as cellular networks or WiFi in this educational context.

Attendance control via Bluetooth has already been implemented in SE \cite{puckdeevongs2020classroom}. Despite BLE provides the benefit of broadcasting mode without the requirement for pairing, this characteristic makes BLE vulnerable to passive sniffing attacks \cite{barua2022security}. Additionally, BLE has limited coverage; therefore, the most feasible approach for user localisation is fingerprinting, as demonstrated in \cite{puckdeevongs2020classroom}. However, fingerprinting is a non-scalable technique that necessitates a comprehensive measurement offline phase.

\subsection{Internet of Things (IoT)}

A wide range of sensors (e.g. smoke detectors, temperature, and proximity sensors) or beacons (e.g. WiFi or Bluetooth) fall under the category of IoT devices, which play a crucial role in the educational sector as a fundamental enabler \cite{sutjarittham2019experiences}. The integration of IoT systems and devices enables multiple applications such as resource monitoring or occupancy tracking \cite{azizi2020application}. IoT information is typically centralized into a system that cross-correlates data from various IoT enablers to provide localization information \cite{anagnostopoulos2021challenges}. 

The use of low-cost sensors in IoT allows effective control and monitoring of large areas, contributing to the optimization of spaces and resources.  Nevertheless, IoT devices are vulnerable to security breaches that may compromise confidential information and personal privacy \cite{meneghello2019iot}.
In addition, it is crucial to consider the diversity of the infrastructure across educational institutions, including universities and schools, as well as the resources and advantages afforded by any financial investments made.

The success of a location-based service is largely determined by the chosen educational establishment. It is evident that universities prioritize the allocation of greater resources towards infrastructure technology in comparison to primary or secondary schools. Some universities, such as the University of Malaga, utilise IoT networks that monitor and investigate the effect of vegetation conditions (temperature, humidity, etc.) on students' comfort levels \cite{fortes2019campus}. The University of Zaragoza utilises a spatial and geographic information system to provide ongoing access to the inventory of its facilities and available classrooms \cite{martinez2021internet}. Similarly, the University of Alicante employs a vehicle mobility management system to monitor use of its car parks \cite{macia2021modelling}.
 

\subsection{Computer Vision}

Multiple camera-based applications exist for real-time monitoring of educational facilities, such as libraries, cafeterias or classrooms \cite{olagoke2020literature}. Furthermore, it is possible to determine levels of occupancy in these spaces using the computer vision, allowing for more efficient use of resources \cite{das2020space}. 

Image processing  constitutes the fundamental element of localization and tracking with computer vision. It provides accurate navigational data which correlates both localization and motion information with centimetric precision \cite{morar2020comprehensive}. This technology relies on fixed cameras placed at strategic locations within the infrastructure, such as campuses or educational settings. To implement navigation and tracking through the SE needs a map of the building and a configuration phase that involves marking the positions of stationary cameras on the map \cite{sun2019see}. The algorithms employed constantly update the navigation status of multiple students based on their current foreground state and previous positions \cite{ashiq2022cnn}. By examining changes in the image structure, computer vision objectively identifies the foreground elements through pixel correlation \cite{saponara2021implementing}. Nevertheless, enlarging the monitored areas results in a significant increase in expense, both in terms of effort and infrastructure, in order to uphold a high level of accuracy.

 \section{Privacy Concerns} \label{PRIV} 


A  potential limitation of using localization in SE are the privacy concerns of students. Institutions may need to implement safeguards to ensure that students' location data is not misused \cite{ayres2010locpris}. To address these privacy concerns, educational institutions may need to implement strict policies and procedures around the collection and use of location data, as well as a clear and transparent disclosure of these policies.

The General Data Protection Regulation (GDPR) is an European Union (EU) regulation that governs the protection of personal data. In relation to localization privacy, GDPR establishes several requirements for companies that collect, process, and store location data \cite{regulation2016regulation}. These requirements include transparency, ensuring that only the minimum amount of data is collected, ensuring that data is accurate and up-to-date, implementing appropriate security measures, limiting data retention and giving individuals the right to access, rectify, and erase their location data \cite{linden2020privacy}. Additionally, GDPR requires institutions to appoint a data protection officer if they process or monitor location data on a large scale or if the core activities of their business involve regular and systematic monitoring of individuals \cite{regulation2016regulation}.



\section{Proof of Concept} \label{POC} 
In this section, the PoC is described in full detail. It implements an attendance control system for SE that locates students within a laboratory using cellular technology and WiFi networks, both operating independently and combined. Different techniques for classification and regression are explained, and the different challenges and limitations that this PoC can encounter are defined. In addition, the experiments and the scenario setups are described. Results of the different techniques and technologies are discussed. Finally, a proposal for an architecture that integrates this service within an OpenRAN derived from this PoC is given.

\subsection{Objectives} 
The main objective consists on localizing the students within a specific classroom depending on their radio signal information for an automatic attendance control system in real-time. To achieve this goal, a comparative analysis is conducted between two different methodologies of Random Forest (RF). 

The first approach utilizes a purely classification-based ML model, which determines whether the student is in a laboratory or not based on the provided input. Conversely, the second approach entails a location regression model that estimates the student's position and subsequently classifies whether the student is situated within a laboratory area. Thus, both systems use, as input, the RSSI signal obtained from 5G and WiFi networks working both together and independently. In this case, the system uses RF method that is  straightforward to implement and can accomplish high precision in classification and regression processes. The objectives of this PoC are disclosed as follows:


\begin{itemize}
    \item Performance validation of the automatic attendance control system.
    \item Comparison of classification and regression success rate.
    \item Demonstration of the viability and benefits of the opportunistic fusion for location-based services for SE.
\end{itemize}

\subsection{Methodology} 

ML techniques enable the prediction of whether the user is inside a classroom. Specifically, RF is used in this PoC because it is a simple technique and can accomplish a high location precision, particularly when used with a significant amount of training data \cite{breiman2001random}. This technique combines the predictions from various single models, known as \textit{base models}, to create a final outcome. RFs are a versatile tool for a range of ML tasks such as classification, regression or anomaly detection \cite{guo2018indoor}. RFs are particularly effective in classification and regression tasks as they can effectively merge the predictions of multiple decision trees to provide a final location of the UE \cite{louppe2021understanding}. 

\begin{figure}[!h]
\centering
\includegraphics[width=\columnwidth]{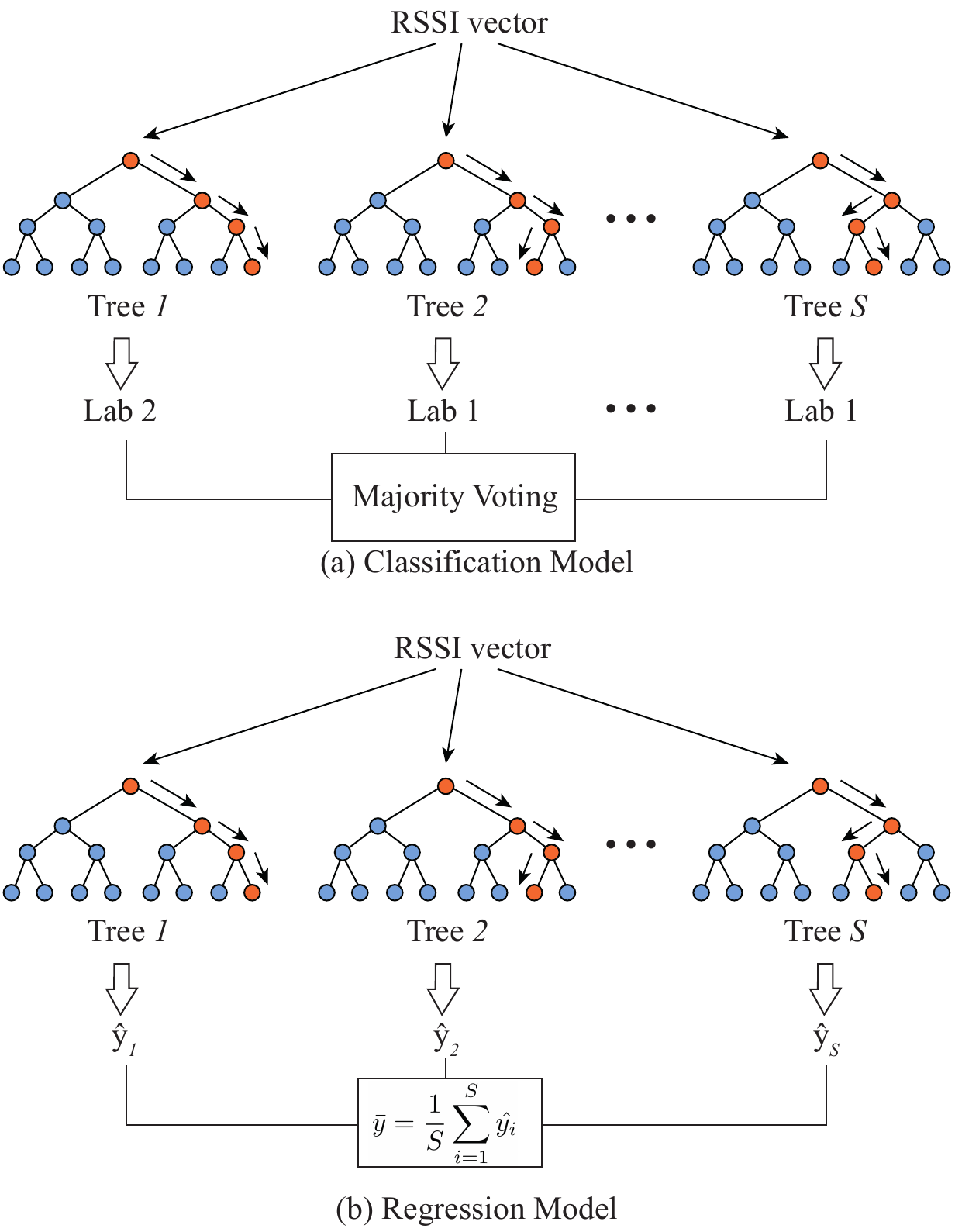}
\caption{RF schema}
\label{RF}
\end{figure}

RFs create decision trees through a process called bootstrapping, which involves randomly selecting a subset of the training data and using this subset to create a decision tree, this process is repeated multiple times, resulting in a large number of decision trees that are all trained on different subsets of the data \cite{cutler2012random}. The final prediction is then made by averaging the predictions of all the decision trees in the forest as illustrated in Figure \ref{RF}. In the case of the classification process, as shown in Figure \ref{RF} (a), a majority voting mechanism is employed to determine the final output, which corresponds to the most commonly voted label. In contrast, in the regression model process, depicted in Figure \ref{RF} (b), the estimation of the location is achieved by averaging the positions across the trees. RFs are robust to the presence of noise in the data, as the averaging process helps to reduce the impact of any individual decision tree that may be making incorrect predictions \cite{roy2012robustness}.

\subsection{Challenges and limitations of the PoC} \label{CHAL_POC} 

Due to the high concentration of students within a classroom environment, the personal devices utilized by these students possess the potential to introduce interference within the radio frequency spectrum degrading the end-user position estimation. In addition, areas without coverage may be generated, relying solely on 5G or WiFi localization for attendance control becomes problematic since students may encounter difficulties in establishing network connections.

In this PoC, the focus is on the basic infrastructure commonly found in primary or secondary schools, including cellular networks and WiFi. As a result, this PoC can be conducted at any educational institue with limited resources.

\subsection{Experimental setup} 

\begin{figure}[!h]
    \centering
  \subfloat{%
       \includegraphics[width=0.75\linewidth]{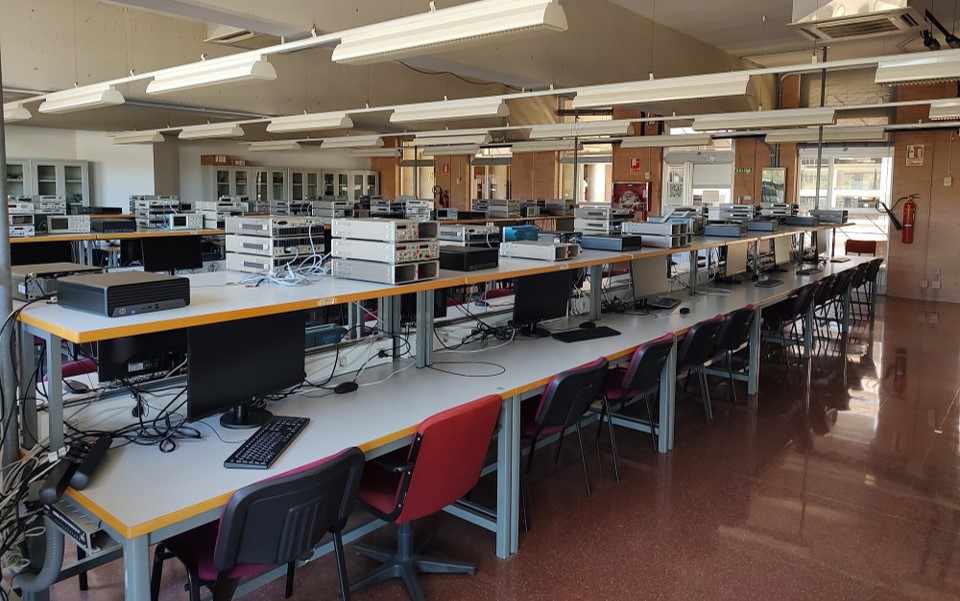}}
    \\
  \subfloat{%
        \includegraphics[width=0.75\linewidth]{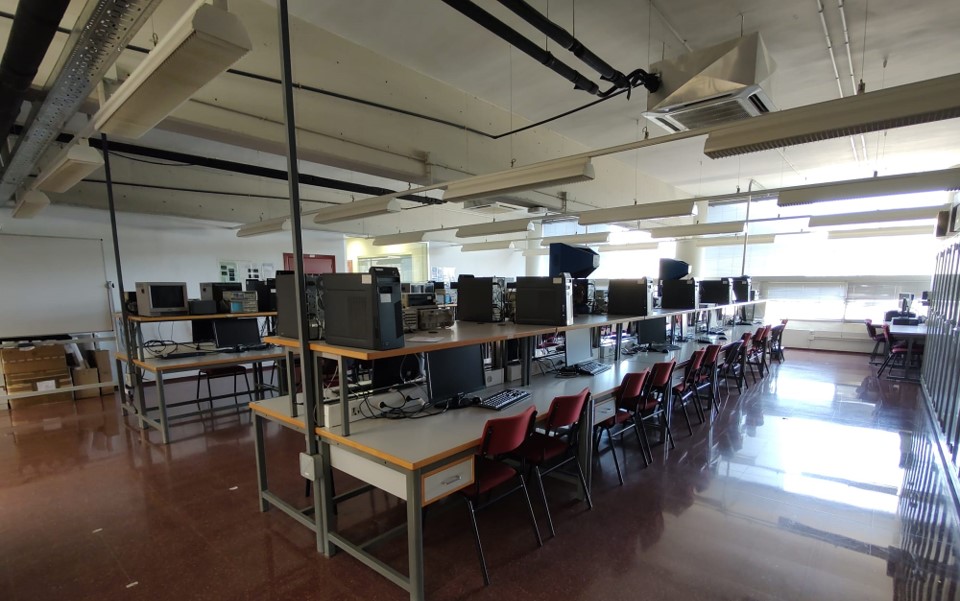}}
    \hfill
  \caption{Images of both laboratories}
  \label{FotoLaboratorios} 
\end{figure}

The scenario where the PoC was deployed is located at the University of Malaga and composed by two different laboratories as shown in Figure \ref{FotoLaboratorios}. It is a medium-cluttered scenario with instrumentation equipment that create signal reflections in the whole area. Both 5G and WiFi networks were utilized to conduct the measurement campaign. 

\begin{figure}[!h]
\centering
\includegraphics[width=\columnwidth]{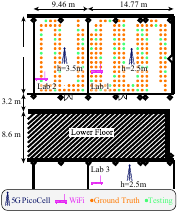}
\caption{Map of the scenario}
\label{MapLaboratory}
\end{figure}

To minimize interference with commercial networks, three gNodeBs (gNBs) were configured for the 5G network, with the stations located at different heights (2.5m and 3.5m), as shown in the map of the scenario map in Figure \ref{MapLaboratory}. Additionally, Google WiFi mesh routers were used as WiFi APs, placed on shelves at a height of 2 meters. For the 5G network, the gNBs were placed in the ceiling to provide good visibility and transmit at a frequency of 3774.990 MHz with a power of 20 dBm. The scenario consisted of two laboratories, with an additional laboratory where there is a 5G AP, which included metallic elements that could potentially cause signal blocking, attenuation, and multipath effects. In this study, a dataset comprising more than 250 samples was employed. To facilitate RF training and evaluation, the data was divided into a training set and a testing set, visually depicted in Figure \ref{MapLaboratory} as orange and green dots, respectively. The 20\% of the measurements were designated for testing. To ensure reliable statistical outcomes, the experiment was repeated a thousand times, with each iteration involving random selection of training and testing points using the Monte Carlo method. Figure \ref{MapLaboratory} illustrates one instance of this iterative process, ensuring precise statistical results. A Motorola Edge 20 smartphone, operating on Android 11, serves as the target UE for determining location. 


To collect the RSSI of serving and neighboring cells in both 5G and WiFi networks, an smartphone application has been created. The data collected by this application is subsequently transmitted via the 5G network to a server, where the measurement samples are stored in a MySQL database for future analysis. The application also includes a feature for indicating the ground truth location, which is then included along with the measured data. By using this setup and collecting data application from real-world scenarios, we aimed to accurately evaluate the performance of the classification system.


\subsection{Results} \label{RES} 
This section presents the performance of the localization and classification results achieved from the PoC, which aimed to assess the reliability of the system for classifying a student's location within a specific classroom. To achieve this goal, the performance of both RF models are evaluated by comparing the accuracy of final classification obtained from 5G and WiFi networks, both together and independently.

\subsubsection{Comparing Accuracy of Classification and Localization-Based Regression Models} 
\begin{figure}[!h]
\centering
\includegraphics[width=\columnwidth]{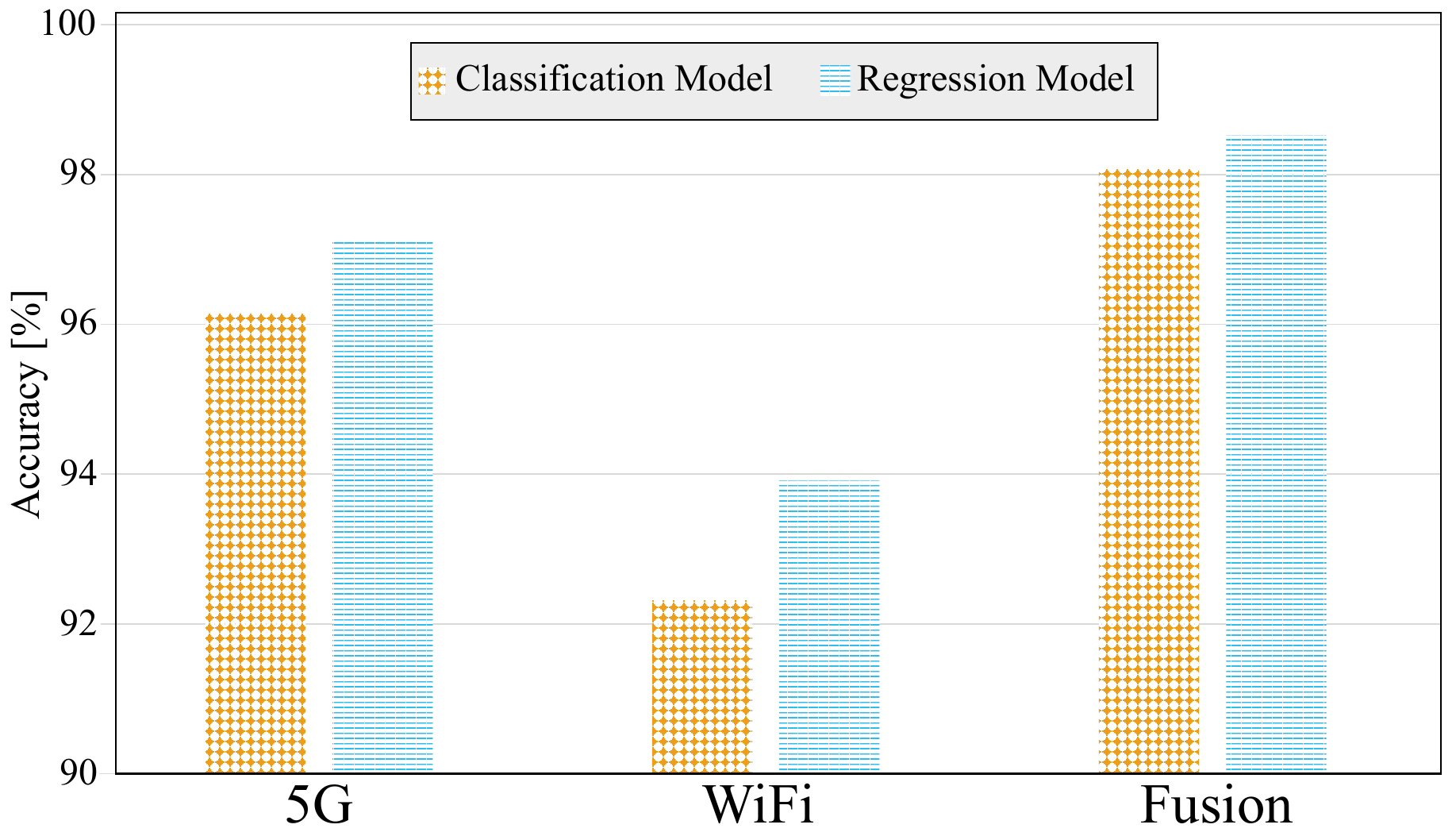}
\caption{Comparative accuracy analysis of classification (orange) and regression (blue) RF models for 5G, WiFi and Fusion}
\label{GraphComparison}
\end{figure}

in order to assess the effectiveness of the classification and localization-based regression models, it is crucial to compare the accuracy of the classification process in correctly identifying the laboratory where the student is placed. Figure \ref{GraphComparison} illustrates the percentage of accuracy performance of the classification (orange) and localization-based regression (blue) models for 5G, WiFi, and their fusion.

The findings demonstrate that the regression model and the fusion of different technologies significantly enhance the overall classification performance of the system. In indoor environments, signal propagation conditions tend to be challenging, resulting in a random reception of RSSI by the UEs due to blockages and multipath effects. Based on the user's position estimation, the system determines whether the user is situated within a classroom. Consequently, in regions closer to the classroom boundaries, there is an enhancement in performance compared to the pure classification method. This enhancement is attributed to the fact that regression systems first estimate the UEs' location and subsequently assign it to a particular classroom.

\subsubsection{Localization performance by regression} 
\begin{figure}[!h]
\centering
\includegraphics[width=\columnwidth]{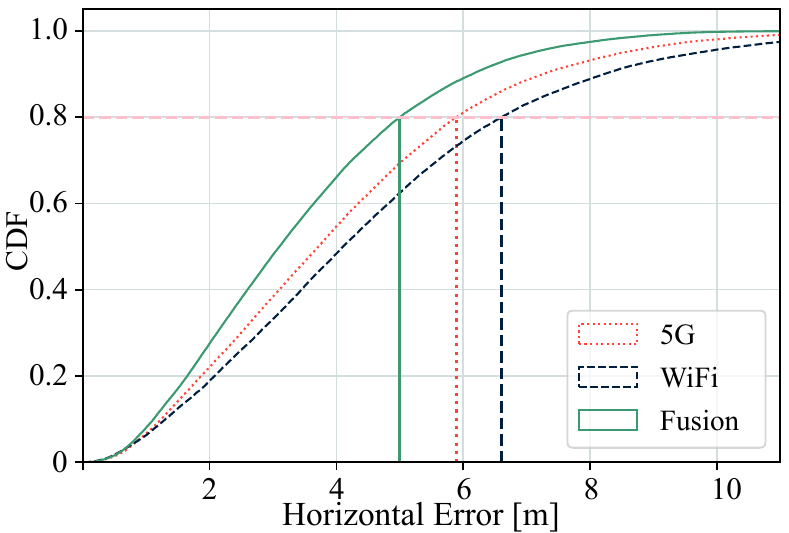}
\caption{CDF of localization regression for 5G, WiFi and fusion}
\label{CdfLocalization}
\end{figure}

the system classifies the student's location in a specific laboratory. Figure \ref{CdfLocalization} represents the Cumulative Density Function (CDF) of the horizontal localization error. The localization error is relative to the disparity between the estimated final position of the user and the ground truth position. As it can be observed, the performance of 5G (red dotted line) is better than WiFi (blue dashed line). However, the fusion (green line) of both technologies improves the overall performance compared with each technology in isolation. Fusing 5G and WiFi achieves a location error of 5 meters in 80\% of the cases, thereby satisfying the location requirements stated in Table \ref{TableComparison} as originally mentioned.


The better precision of the regression, the better performance at the classification process is expected. Notably, when examining the data independently, 5G obtains higher localization precision resulting in a higher accuracy classification results. Consequently, when combining 5G with WiFi data, there is a noticeable enhancement in classification accuracy than in isolation. 

By combining two technologies, the system not only improves its classification accuracy but also extends its coverage area beyond what could be achieved with a single technology. Simultaneously employing two supplementary technologies offers redundancy in case of system failure or outage, further enhancing the robustness of the system.



\section{Conclusion} \label{CONC} 

Location-based services are increasingly used in education to enhance the learning experience and increase efficiency, with the implementation of XR and indoor navigation on campuses. 

This paper aims to provide an overview of localization in SE with real-world examples, analyzing the main technologies and techniques employed to improve the quality of education. After analyzing the challenges and limitations of the different technologies in this educational context, we conducted an experiment on a system created to automate attendance control in education settings with a limited budget. The system effectively demonstrated the advantages of combining various available technologies within educational institutions, as shown by real data. 

This PoC illustrates that a localization-based regression model performs better than a simple classifier model. The proposed AAC system can be readily implemented in academic settings, offering a straightforward and unobtrusive method for enhancing teaching and learning efficiency. The attendance monitoring procedure may be executed using students' mobile devices, as all such devices come equipped with both cellular and WiFi technologies. Additionally, the system's data can be analyzed to identify attendance patterns, allowing teachers to optimize class scheduling and delivery. The proposed system streamlines attendance tracking and could provide other location-based services for students, including space usage and occupancy tracking. The potential of wide-spread 5G and WiFi technologies in the education sector to revolutionise how students learn and interact with their surroundings is significant. 


\bibliographystyle{IEEEtran}
\bibliography{main}


\begin{IEEEbiography}[{\includegraphics[width=1in,height=1.25in,clip,keepaspectratio]{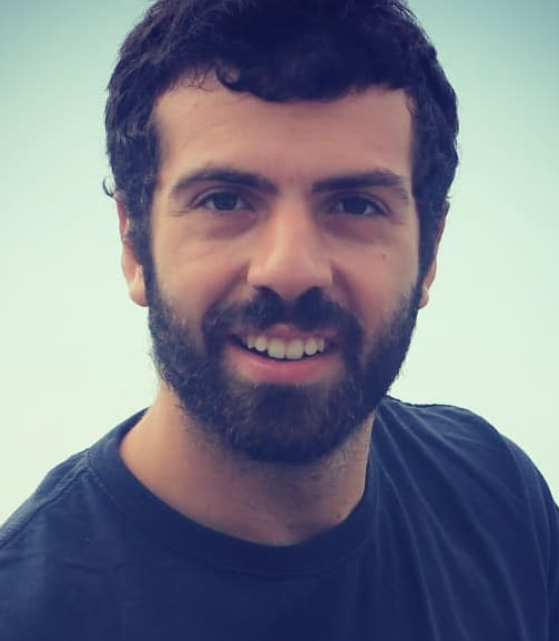}}]{Carlos Simón Álvarez Merino}
received his B.Sc. degree in Telecommunication Technologies and his dual M.Sc. degree in Telecommunication Engineering and Telematics and Networks from the University of Málaga, Spain, in 2017 and 2019, respectively. He is currently working as a Research Assistant at the Communications Engineering Department meanwhile he is currently pursuing a Ph.D. degree.
\end{IEEEbiography}
\vskip -2\baselineskip plus -1fil

\begin{IEEEbiography}[{\includegraphics[width=1in,height=1.25in,clip,keepaspectratio]{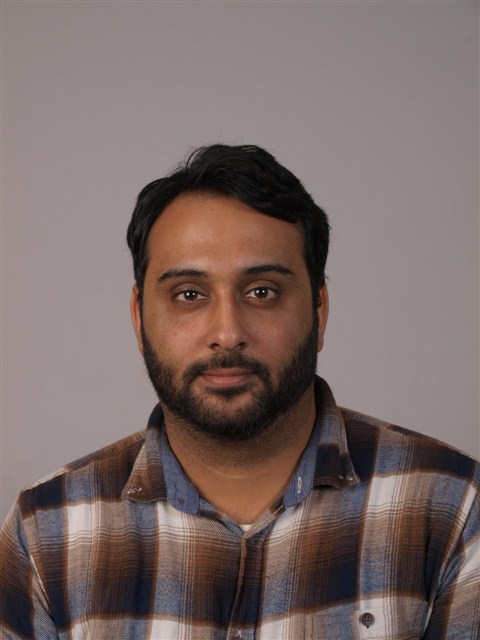}}]{Emil Jatib Khatib}
My main research interests are Machine Learning and Big Data Analytics applied to 5G, IoT and Industry 4.0 applications. I enjoy seeing my system designs working in reality, so I have actively worked on testbeds in the Mobilenet group (where I worked on AI/ML for network management and worked on the development of an HTTP/REST API as an enabler of SON algorithms) and with our collaborators in Aalborg University (where I developed a multi-connectivity testbed using USRPs and a prediction scheme for URLLC in in industrial scenarios). Currently, I am working on assessing the performance of wireless connections in an assembly line, using the ML toolset provided by our team.
\end{IEEEbiography}
\vskip -2\baselineskip plus -1fil

\begin{IEEEbiography}[{\includegraphics[width=1in,height=1.25in,clip,keepaspectratio]{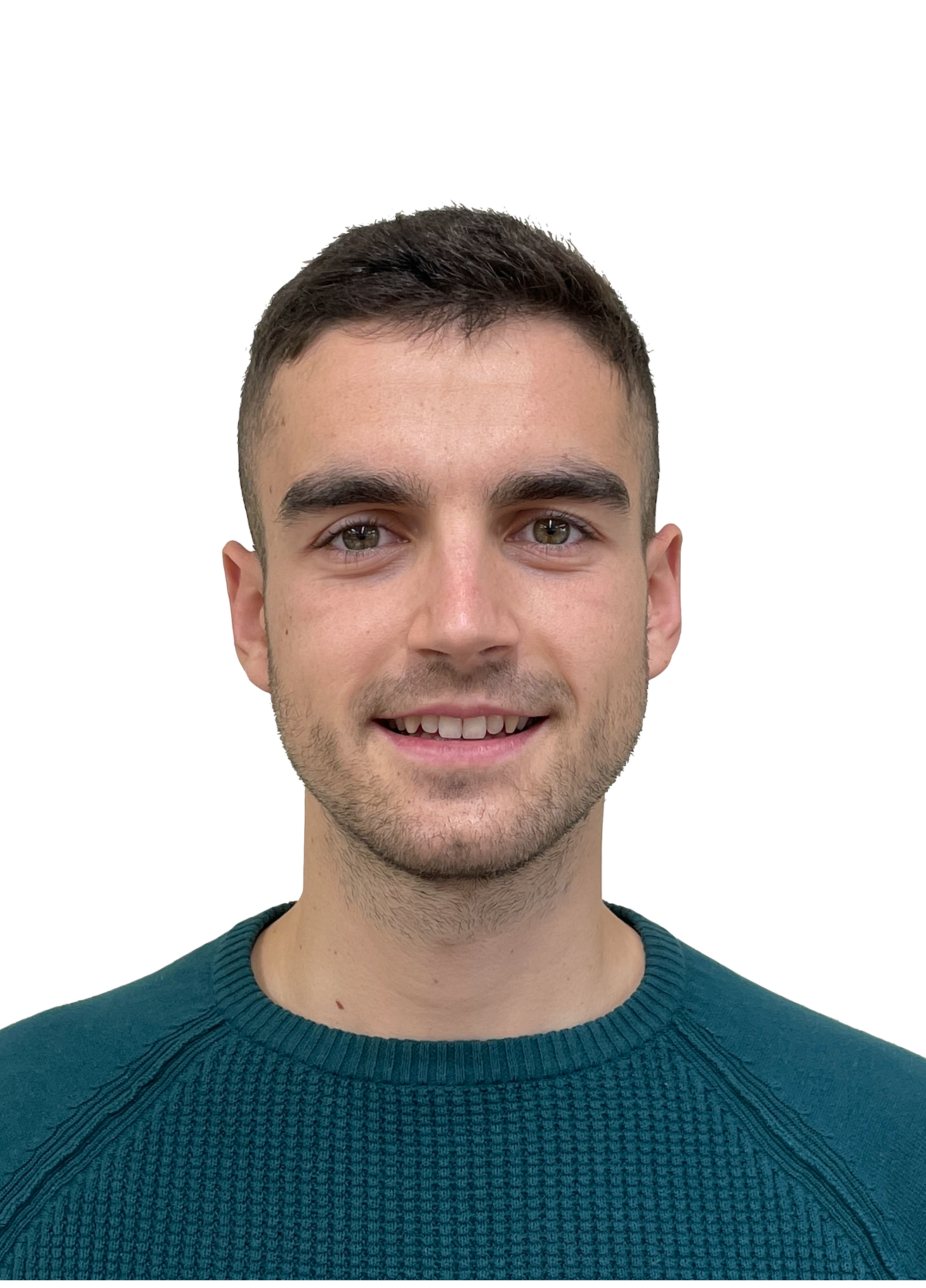}}]{Antonio Tarrías Muñoz}
received a M.S. in Telematics and Telecommunication Networks in 2020 from the University of Malaga, Spain. In 2019, he started to work as research assistant and assistant professor with the Department of Communications Engineering. Currently, he is pursuing the Ph.D. degree in the field of cellular communications.
\end{IEEEbiography}
\vskip -2\baselineskip plus -1fil


\begin{IEEEbiography}[{\includegraphics[width=1in,height=1.25in,clip,keepaspectratio]{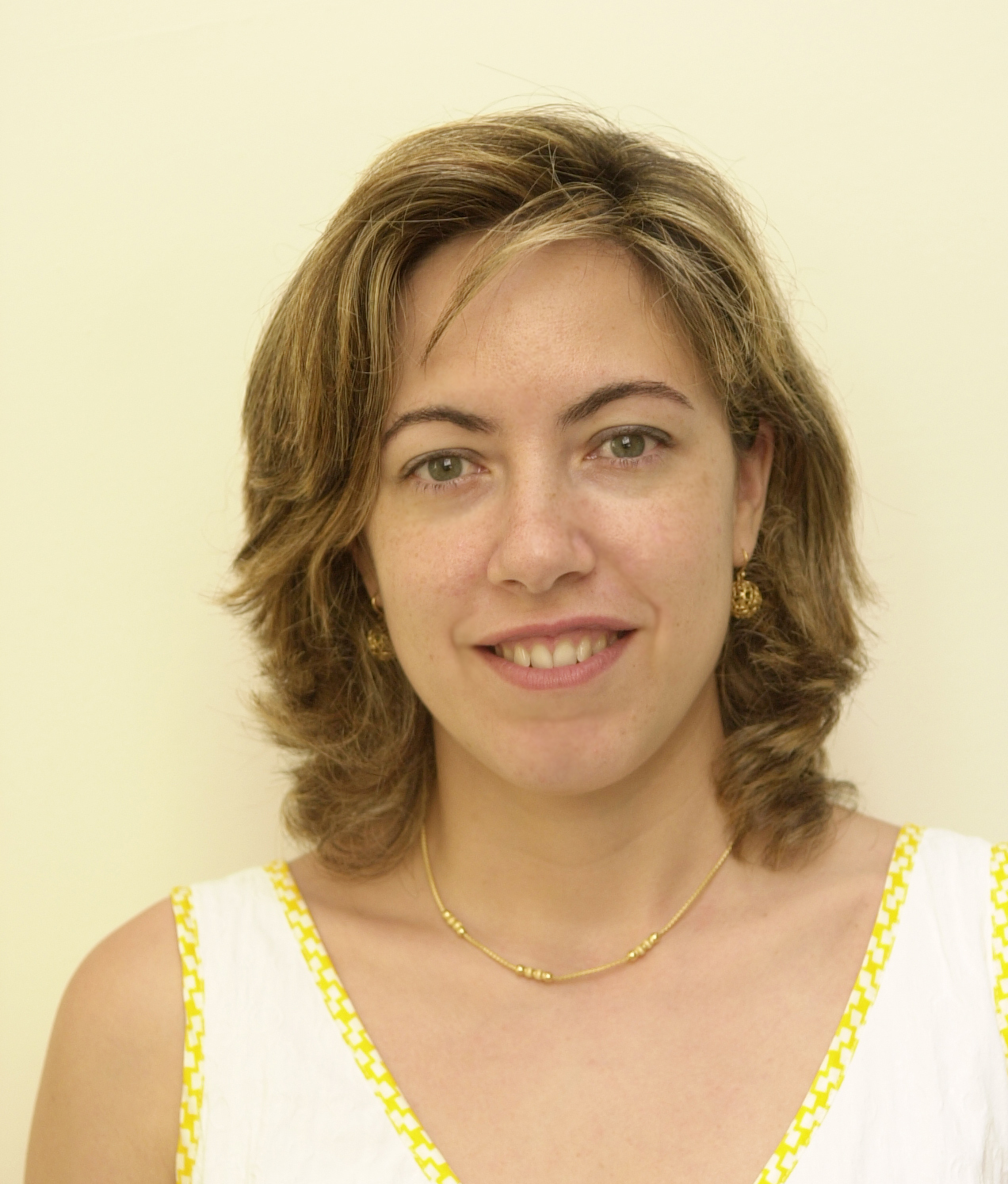}}]{Raquel Barco Moreno}
holds a M.Sc. and a Ph.D. in Telecommunication Engineering from the University of Malaga. From 1997 to 2000, she worked at Telefónica in Madrid (Spain) and at the European Space Agency (ESA) in Darmstadt (Germany). In 2000, she joined the University of Malaga, where she is currently Full Professor. She took part as researcher in a Nokia Competence Center on Mobile Communications for three years. She has led projects with the main mobile communications operators and vendors for a value $>$ 7 million €, she is author of 7 patents and has published more than 100 high impact journals and conferences.
\end{IEEEbiography}
\vskip -2\baselineskip plus -1fil

\end{document}